\documentclass[conference]{IEEEtran}
\IEEEoverridecommandlockouts

\usepackage{cite}
\usepackage{amsmath,amssymb,amsfonts}
\usepackage{algorithm}
\usepackage{algpseudocode}
\usepackage{graphicx}
\usepackage{booktabs}
\usepackage{xcolor}
\usepackage{tikz}
\usepackage{pgfplots}
\pgfplotsset{compat=1.18}
\usetikzlibrary{arrows.meta,positioning,shapes.geometric,calc,fit}
\usepackage{url}
\usepackage{listings}
\usepackage[hidelinks]{hyperref}

\newcommand{\code}[1]{\texttt{#1}}

\begin{document}

\title{AuthProbe: Specification-Driven, Multi-Identity\\
Detection of Broken Object-Level Authorization\\ in Recruitment APIs}

\author{\IEEEauthorblockN{Jay Barach}
\IEEEauthorblockA{\textit{Independent Researcher}\\
\url{https://github.com/jbarach2012/AuthProbe}}}

\maketitle

\begin{abstract}
Broken Object-Level Authorization (BOLA), also known as Insecure Direct Object
Reference (IDOR), has topped the OWASP API Security ranking since 2019 and is the
root cause of some of the largest exposures of applicant data in recruitment
technology. The defining feature of this flaw class is that a malicious request
is byte-for-byte indistinguishable from a legitimate one, which is precisely why
web application firewalls and single-identity scanners fail to catch it. We
present AuthProbe, an open-source, black-box scanner that detects BOLA and IDOR
in HTTP APIs by driving its tests from an OpenAPI specification and by acting
under two or more identities that the operator controls. AuthProbe discovers,
for each identity, the objects that identity legitimately owns, then attempts to
read one identity's objects while authenticated as another and confirms a leak by
comparing the response against a ground-truth fetch by the true owner. It also
walks predictable identifiers to expose enumeration and reports missing
authentication and existence oracles. The tool returns a severity-thresholded
exit code and machine-readable reports so that it can gate a continuous
integration build. On a synthetic recruitment API in which the McHire failure
class is reproduced, AuthProbe detects every planted cross-identity read with no
false positives on a hardened counterpart, and its running time grows linearly
with the number of objects under test. AuthProbe is released under the Apache 2.0
license with an authorized-use guardrail.
\end{abstract}


\section{Introduction}
Automated hiring platforms now sit between most job seekers and most employers.
An applicant tracking system concentrates an unusually rich store of personal
data, including names, contact details, home addresses, employment history,
personality assessment results, and full chat transcripts, while it is optimized
for speed, automation, and scale. That combination has repeatedly come at the
expense of security fundamentals, and the fundamental most often neglected is
authorization at the level of the individual object.

The clearest recent illustration is the McHire incident of 2025. Two
researchers logged into a dormant administrative test account on a widely used
recruitment chatbot platform using the credentials \code{123456} for both the
user name and the password, with no second factor required. They then found an
internal endpoint that accepted a sequential applicant identifier. By decrementing
that number they could retrieve any applicant's chat transcript and contact
details, and the exposure spanned on the order of sixty four million records
\cite{mchire}. The platform vendor disputes the real-world scale and states that
only a handful of records were viewed by the researchers, yet the potential
exposure and the underlying flaw are not in dispute. The flaw is a textbook case
of Broken Object-Level Authorization: the endpoint authenticated the caller but
never checked whether that caller was permitted to read the specific object being
requested.

BOLA has held the top position in the OWASP API Security Top Ten since the list
was first published in 2019, and it retains that position in the 2023 revision as
API1:2023 \cite{owaspapi}. Industry telemetry places it among the most frequently
exploited API weaknesses \cite{salt,imperva}. The reason this class is so
persistent is structural. A BOLA request uses a valid session and a
well-formed path, so it carries no payload signature, no anomalous syntax, and no
injection string. A signature-based web application firewall sees a normal
request, and a conventional dynamic scanner that operates under a single identity
has no notion of which objects belong to which principal, so it cannot recognize
that a returned object was one the caller should not have seen. Detecting BOLA
therefore requires contextual testing that is aware of ownership and that operates
under more than one identity at once.

This paper presents AuthProbe, a scanner built specifically for that task. Its
design rests on three observations. First, modern APIs publish a machine-readable
contract in the form of an OpenAPI document, which enumerates the endpoints and
identifies which ones return individual objects. Second, an authorization flaw is
only demonstrable when the tester holds at least two identities and can show that
one reaches the other's data. Third, a useful security check must fit into an
automated pipeline, which means it must produce a deterministic verdict and a
machine-readable report rather than a human-oriented narrative. AuthProbe unites
these three observations into a single tool.

The scope of this work is deliberate. We target the read path of the item
operation, which is the dominant Broken Object-Level Authorization pattern and the
one behind the motivating incident, and we optimize for a low false-positive rate
so that the tool can run unattended on every build rather than as an occasional,
expert-supervised audit. We do not attempt to model every access-control property,
and we treat source-level analysis as a complement rather than a competitor. This
focus is what lets AuthProbe be both simple to deploy and trustworthy enough to
gate a release.

We make the following contributions.
\begin{itemize}
\item A black-box detection method for BOLA and IDOR that is driven by an OpenAPI
specification, requires no access to the target's source code, and confirms a
finding by response differencing against a ground-truth fetch, which keeps the
false-positive rate low (Sections \ref{sec:threat} and \ref{sec:method}).
\item A set of complementary probes for identifier enumeration, missing
authentication, and object-existence oracles, together with a severity model and
a continuous integration gate (Sections \ref{sec:method} and \ref{sec:impl}).
\item An open-source implementation that ships with an intentionally vulnerable
recruitment API and a hardened counterpart, so the tool can be demonstrated and
validated end to end without touching a third-party system (Section
\ref{sec:impl}).
\item An evaluation on that testbed showing complete detection of the planted
McHire-class flaw, zero findings on the hardened target, and running time that
scales linearly with the number of objects (Section \ref{sec:eval}).
\end{itemize}
AuthProbe is a defensive tool. It is intended for use only against systems the
operator owns or is authorized to test, and it enforces that intent with a
runtime guardrail described in Section \ref{sec:ethics}.

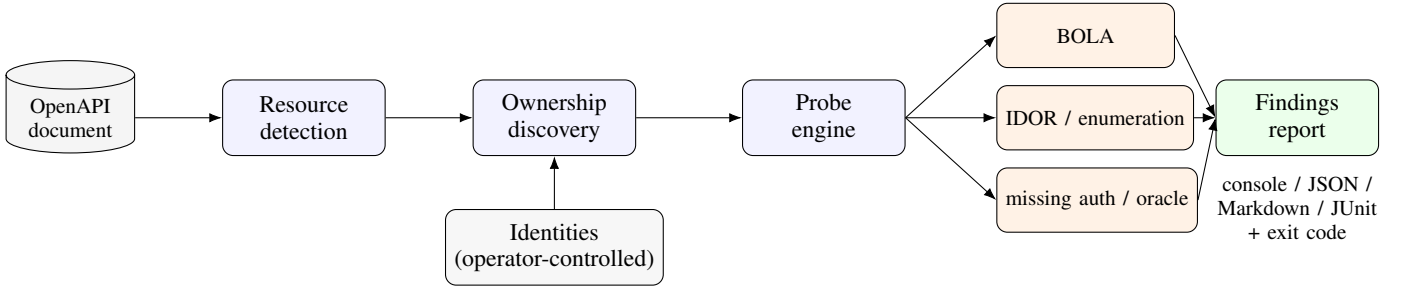
\begin{figure*}[t]
\centering
\begin{tikzpicture}[
  >=Latex, node distance=0.7cm and 1.15cm,
  box/.style={draw, rounded corners, align=center, minimum height=1.0cm,
              minimum width=2.15cm, fill=blue!5, font=\small},
  db/.style={draw, cylinder, shape border rotate=90, aspect=0.28, align=center,
             fill=gray!8, minimum height=1.1cm, minimum width=1.7cm, font=\footnotesize},
  probe/.style={draw, rounded corners, align=center, minimum height=0.85cm,
                minimum width=2.35cm, fill=orange!10, font=\footnotesize}
]
\node[db] (spec) {OpenAPI\\ document};
\node[box, right=of spec] (detect) {Resource\\ detection};
\node[box, right=of detect] (disc) {Ownership\\ discovery};
\node[box, right=1.4cm of disc] (engine) {Probe\\ engine};
\node[probe, above right=0.15cm and 1.2cm of engine] (p1) {BOLA};
\node[probe, right=1.2cm of engine] (p2) {IDOR / enumeration};
\node[probe, below right=0.15cm and 1.2cm of engine] (p3) {missing auth / oracle};
\node[box, right=4.1cm of engine, fill=green!8] (report) {Findings\\ report};
\node[box, below=0.7cm of disc, fill=gray!6] (ids) {Identities\\ (operator-controlled)};

\draw[->] (spec)--(detect);
\draw[->] (detect)--(disc);
\draw[->] (ids)--(disc);
\draw[->] (disc)--(engine);
\draw[->] (engine.east)-- (p1.west);
\draw[->] (engine.east)-- (p2.west);
\draw[->] (engine.east)-- (p3.west);
\draw[->] (p1.east)-- (report.west);
\draw[->] (p2.east)-- (report.west);
\draw[->] (p3.east)-- (report.west);
\node[below=0.15cm of report, font=\footnotesize, align=center] {console / JSON /\\ Markdown / JUnit\\ + exit code};
\end{tikzpicture}
\caption{AuthProbe pipeline. The OpenAPI document yields the set of
object-returning resources. For each resource the tool discovers which objects
each operator-controlled identity owns, then runs four probes and emits a report
plus a severity-thresholded exit code suitable for a continuous integration gate.}
\label{fig:arch}
\end{figure*}

\section{Background and Related Work}
\label{sec:related}

\subsection{Broken object-level authorization}
Access control on a resource involves two distinct questions. Authentication asks
who the caller is, and authorization asks whether that caller may perform the
requested action on the requested object. BOLA is the failure of the second
question at the granularity of an individual object. The OWASP API Security
project defines it as an API endpoint that receives an object identifier from the
client and acts on it without verifying that the authenticated caller is entitled
to that specific object \cite{owaspapi}. The corresponding weakness in the Common
Weakness Enumeration is CWE-639, authorization bypass through a user-controlled
key, which sits under the broader CWE-284, improper access control \cite{cwe639}.
IDOR is the common name for the same defect when the user-controlled key is a
direct reference such as a database row identifier.

\subsection{Access control testing}
Detecting authorization flaws has an established research literature. AuthScope
drives a mobile application automatically, learns the request fields that carry
object identifiers through differential traffic analysis, and substitutes one
user's identifier into another user's session to reveal missing checks
\cite{authscope}. MACE analyzes web applications to find privilege escalation
paths that arise from inconsistent access-control enforcement \cite{mace}.
FlowWatcher defends against data disclosure by tracking an application's intended
ownership policy and blocking responses that violate it \cite{flowwatcher}. More
recently, BolaRay studies real BOLA vulnerabilities in database-backed
applications, distills four recurring object-level authorization models, and
combines SQL analysis with static analysis to check whether the code enforces the
appropriate model \cite{bolaray}. These works are powerful, and several achieve
high precision, but most require either the application binary, the source code,
or instrumentation of the running server. AuthProbe occupies a different and
complementary point in the design space. It is fully black box, it needs only the
published specification and network access under identities the operator holds,
and it is built to run unattended inside a delivery pipeline. It does not compete
with a static analyzer such as BolaRay so much as it covers the deployment-time
gap that static analysis cannot reach, namely a running service whose source may
be unavailable.

\subsection{Specification-driven API testing}
A separate line of work generates tests from an API contract. RESTler infers
producer and consumer dependencies among operations and performs stateful fuzzing
of REST services \cite{restler}. RESTest applies black-box, constraint-based
testing driven by the specification \cite{restest}, and EvoMaster uses
evolutionary search to generate test cases for RESTful APIs \cite{evomaster}.
These tools target functional faults and server errors rather than authorization
policy, and they typically operate under a single identity. AuthProbe borrows the
idea of treating the specification as the source of truth for what to test, then
redirects it toward a security property that only becomes visible when the tester
holds multiple identities. General-purpose dynamic scanners such as the OWASP Zed
Attack Proxy are widely used to crawl and fuzz web applications \cite{zap}, and
they can be scripted to compare responses across sessions, but they do not model
object ownership out of the box and are not driven by an object-level authorization
contract, so a team must build the multi-identity comparison itself. AuthProbe
packages exactly that comparison as a first-class, specification-driven check.

\subsection{Authorization systems}
On the defensive side, relationship-based authorization systems such as Google
Zanzibar provide a consistent, centralized way to answer object-level access
questions at scale \cite{zanzibar}, and its open reimplementations have become the
recommended way to enforce the checks whose absence AuthProbe detects. Vulnerable
application benchmarks built around the OWASP API risks, such as the one described
by Idris and colleagues \cite{idrisvuln}, motivate the value of a shipped,
intentionally vulnerable target for validating a detector. AuthProbe complements
these by providing the offensive test that confirms whether a given deployment
actually enforces the policy that a system like Zanzibar would express.

\subsection{Why recruitment APIs are a distinctive target}
Recruitment platforms combine three properties that make object-level
authorization both critical and frequently mishandled. They concentrate a broad
range of personal data for every applicant, they are integration dense because an
applicant tracking system typically connects to chatbots, background-check
services, job boards, and analytics through additional APIs, and they are built
under commercial pressure to onboard applicants quickly and at scale. The result
is a large, high-value object store reached through many endpoints, where a single
missing ownership check can expose the entire applicant population. The motivating
incident followed exactly this shape: a chatbot front end collected applicant data,
an internal API returned individual records by identifier, and the absence of an
object-level check turned a predictable identifier into a corpus-wide leak
\cite{mchire}. A detector aimed at this domain must therefore treat the individual
applicant record as the unit of protection and must reason about who owns each
record, which is the design center of AuthProbe.

\subsection{Positioning}
Table \ref{tab:compare} places AuthProbe among representative prior approaches.
The functional testing tools generate load and inputs from a specification but do
not model ownership, so they surface server errors rather than authorization
leaks. The access-control analyzers model authorization well but generally need
the source, the binary, or server instrumentation. AuthProbe is the combination
that the deployment stage requires: black box, specification driven, aware of
multiple identities, and built to gate a pipeline. None of the prior systems
occupies all four positions at once, and it is the union of these properties,
rather than any single one, that makes a check both trustworthy and cheap enough to
run continuously. The comparison is not a claim of superiority on detection depth,
where source-level analyzers retain an advantage, but a statement about where in
the software lifecycle each approach is usable.

\begin{table}[t]
\caption{AuthProbe compared with representative prior approaches.}
\label{tab:compare}
\centering
\footnotesize
\begin{tabular}{@{}lccccc@{}}
\toprule
Approach & Black & No src & Spec & Multi & CI \\
 & box & needed & driven & identity & native \\
\midrule
AuthScope \cite{authscope}      & Yes & Yes & No  & Yes & No  \\
MACE \cite{mace}                & No  & No  & No  & Yes & No  \\
BolaRay \cite{bolaray}          & No  & No  & No  & N/A & No  \\
FlowWatcher \cite{flowwatcher}  & No  & No  & No  & Yes & No  \\
RESTler \cite{restler}          & Yes & Yes & Yes & No  & Part \\
RESTest \cite{restest}          & Yes & Yes & Yes & No  & Part \\
EvoMaster \cite{evomaster}      & Part& No  & Yes & No  & Part \\
\textbf{AuthProbe}              & \textbf{Yes} & \textbf{Yes} & \textbf{Yes} & \textbf{Yes} & \textbf{Yes} \\
\bottomrule
\end{tabular}
\end{table}

\section{Threat Model and Problem Definition}
\label{sec:threat}
We consider an HTTP API that manages objects on behalf of principals. Let $U$ be
the set of principals and let $R$ be a resource type, for example an application
record. Each object $o$ of type $R$ has an owner $\mathrm{own}(o) \in U$ and is
addressed through an identifier. The API exposes a collection operation that lists
the objects visible to the caller and an item operation that returns a single
object given its identifier.

We define ownership as the ground truth that the API itself asserts through its
collection operation. If principal $u$ lists the collection and the object $o$
appears, then $u$ is entitled to read $o$. This is a conservative and
self-consistent definition because it is derived from the target's own behavior
rather than from an external policy that the tester would otherwise have to guess.

The adversary in our model is a principal $a \in U$ who holds a valid session but
attempts to read an object $o$ with $\mathrm{own}(o) \neq a$ and $o$ not in the
set that $a$ is entitled to. A Broken Object-Level Authorization vulnerability
exists for the item operation if there is a principal $a$ and an object $o$ such
that $a$ is not entitled to $o$ yet the item operation, invoked with $a$'s
credentials and the identifier of $o$, returns $o$.

The detection problem is to decide, for a given API and a given set of
tester-controlled identities, whether such a pair $(a, o)$ exists, and to do so
from outside the server with no view of its internals. Two subtleties shape the
method. First, a response with a success status is not sufficient evidence of a
leak, since some APIs return a generic body for a denied request; the tool must
confirm that the returned object is in fact the victim's object. Second, the
tester can only reason about objects it can name, so it must first learn which
identifiers exist and who owns them, which is the purpose of the ownership
discovery step.

\section{System Architecture}
\label{sec:arch}
Figure \ref{fig:arch} shows the AuthProbe pipeline. The input is a target base
URL, an OpenAPI document that is either fetched from the running service or
supplied as a file, and a configuration that lists two or more identities the
operator controls. Each identity is expressed as a set of HTTP headers, which
accommodates bearer tokens, API keys, and session cookies without special cases.

The first stage detects resources. AuthProbe scans the specification for a
collection path and a sibling item path that carries a single path parameter, for
example a list at \code{/applications} beside a fetch at
\code{/applications/\{app\_id\}}. Each such pair becomes a resource with a list
path, a fetch path, a path parameter name, and an identifier field that the tool
infers from the response schema and defaults to \code{id}. The operator may
override this detection in configuration when a contract does not follow the
common convention.

The second stage discovers ownership. For every identity and every resource,
AuthProbe calls the collection operation and records the identifiers that the
operation returns for that identity. This yields, for each identity, the set of
objects the target itself considers visible to it, which is exactly the ground
truth the threat model relies upon.

The third stage runs the probes. The engine executes four probes per resource,
described in Section \ref{sec:method}, and each probe emits zero or more findings.
The final stage renders the findings to the console and, on request, to JSON,
Markdown, and JUnit formats, and it sets the process exit code according to a
configured severity threshold so that a pipeline step fails when a finding at or
above that threshold is present.

The tool is deliberately stateless between runs and requires no agent inside the
target. This keeps deployment simple: a continuous integration job starts the
service, runs AuthProbe against it, and inspects the exit code, in the same way it
would run a unit test suite.

\section{Detection Methodology and Algorithms}
\label{sec:method}

\subsection{Resource detection}
Algorithm \ref{alg:detect} formalizes resource detection. The tool walks the
paths in the specification, matches each path against the pattern of a collection
followed by a single brace-delimited parameter, and pairs it with the
corresponding collection path when both expose a read operation.

\begin{algorithm}[t]
\caption{Specification-driven resource detection}
\label{alg:detect}
\begin{algorithmic}[1]
\Require OpenAPI document $S$
\Ensure set of resources $\mathcal{R}$
\State $\mathcal{R} \gets \emptyset$
\For{each path $p$ in $S.\text{paths}$}
  \If{$p$ matches \code{(coll)/\{param\}} and $p$ has GET}
    \State $c \gets \text{coll}(p)$
    \If{$c \in S.\text{paths}$ and $c$ has GET}
      \State $f \gets \text{inferIdField}(S, p)$
      \State $\mathcal{R} \gets \mathcal{R} \cup \{(\text{name}(c), c, p, \text{param}(p), f)\}$
    \EndIf
  \EndIf
\EndFor
\State \Return $\mathcal{R}$
\end{algorithmic}
\end{algorithm}

\subsection{Ownership discovery and BOLA detection}
The core of the method is Algorithm \ref{alg:bola}. For each resource, the tool
first records the objects each identity owns by reading the collection operation.
It then fetches every owned object as its true owner to capture a ground-truth
view. Finally, for every ordered pair of distinct identities it fetches the
victim's objects while authenticated as the attacker and applies the leak
predicate.

The leak predicate is the heart of the false-positive control. For an attacker
$a$ requesting the identifier of a victim object $o$, let the response be $r$. We
declare a leak when the status of $r$ is a success and the body of $r$ is an
object whose identifier field equals the identifier of $o$. Formally,
\begin{equation}
\mathrm{Leak}(a,o) \equiv \mathrm{ok}(r) \wedge \mathrm{obj}(r) \wedge
\big(\mathrm{id}(r) = \mathrm{id}(o)\big),
\end{equation}
and a finding is raised when $\mathrm{Leak}(a,o)$ holds for an object $o$ that
belongs to another identity and does not belong to $a$. Requiring the returned
identifier to match the requested one rules out the common case in which a denied
request returns a generic error body with a success status, and it rules out
responses that echo a placeholder rather than the victim's record.

\begin{algorithm}[t]
\caption{Ownership discovery and BOLA detection}
\label{alg:bola}
\begin{algorithmic}[1]
\Require resource $R$, identities $I$, client $H$
\Ensure findings $F$
\State $F \gets \emptyset$;\quad $\text{own} \gets \{\}$;\quad $\text{view} \gets \{\}$
\For{each identity $u \in I$} \Comment{ownership discovery}
  \State $\text{own}[u] \gets \text{ids}(H.\text{get}(R.\text{list}, u))$
\EndFor
\For{each $u \in I$, each $o \in \text{own}[u]$} \Comment{ground truth}
  \State $\text{view}[o] \gets H.\text{get}(R.\text{fetch}(o), u)$
\EndFor
\For{each attacker $a \in I$}
  \For{each victim $v \in I$ with $v \neq a$}
    \For{each $o \in \text{own}[v]$ with $o \notin \text{own}[a]$}
      \State $r \gets H.\text{get}(R.\text{fetch}(o), a)$
      \If{$\mathrm{ok}(r) \wedge \mathrm{obj}(r) \wedge \mathrm{id}(r)=o$}
        \State add BOLA finding $(a, v, o)$ to $F$
      \EndIf
    \EndFor
  \EndFor
\EndFor
\State \Return $F$
\end{algorithmic}
\end{algorithm}

The cost of Algorithm \ref{alg:bola} is dominated by its final triple loop, which
performs $O(|I|^2 \cdot m)$ item requests, where $m$ is the number of objects per
identity. In practice the number of tester identities is small and fixed, so the
cost is linear in the number of objects, a property we confirm empirically in
Section \ref{sec:eval}.

\subsection{Identifier enumeration}
Algorithm \ref{alg:idor} addresses the enumeration facet of the McHire flaw. It
first flags identifiers as enumerable when every observed identifier is numeric,
which signals a predictable and therefore walkable scheme. It then takes an
identifier the attacker owns and probes a bounded neighborhood around it. Reaching
any object the attacker does not own is a demonstration that the identifiers are
both predictable and unprotected, which is the precise combination that made the
McHire exposure possible.

\begin{algorithm}[t]
\caption{Identifier enumeration probe}
\label{alg:idor}
\begin{algorithmic}[1]
\Require resource $R$, identity $u$, radius $k$, client $H$
\Ensure findings $F$
\State $F \gets \emptyset$
\If{all observed identifiers are numeric}
  \State add enumerable-identifier finding to $F$
\EndIf
\State $b \gets$ an identifier owned by $u$
\For{$\delta \in \{\pm 1, \dots, \pm k\}$}
  \State $c \gets b + \delta$
  \State $r \gets H.\text{get}(R.\text{fetch}(c), u)$
  \If{$\mathrm{ok}(r) \wedge \mathrm{obj}(r) \wedge c \notin \text{own}[u]$}
    \State record $c$ as reached by enumeration
  \EndIf
\EndFor
\If{any $c$ was reached}
  \State add IDOR finding to $F$
\EndIf
\State \Return $F$
\end{algorithmic}
\end{algorithm}

\subsection{Missing authentication and existence oracle}
Two further probes round out coverage. The missing-authentication probe repeats
the collection and item requests with no credentials and raises a critical finding
when data is returned, which corresponds to API2:2023 in the OWASP list. The
existence-oracle probe compares the status returned for an unauthorized but
existing object with the status returned for a non-existent one. When a denied
request yields a forbidden status while a missing object yields a not-found
status, the difference lets an attacker confirm which identifiers are real, and
the tool reports a low-severity finding. A hardened service avoids this by
returning a uniform not-found status in both cases.

\subsection{Severity model and continuous integration gate}
Each finding carries a severity drawn from an ordered scale of informational, low,
medium, high, and critical. Missing authentication is critical, a confirmed
cross-identity read and a successful enumeration walk are high, a purely
enumerable identifier scheme is medium, and an existence oracle is low. The tool
exits with a non-zero status when any finding reaches a configured threshold,
which defaults to high. This turns AuthProbe into a build gate: a change that
reintroduces an object-level authorization flaw causes the pipeline step to fail
before the change reaches production.

\subsection{Complexity, soundness, and completeness}
Let $I$ be the set of tester identities and let $m$ be the number of objects
discovered per identity. Ownership discovery issues $|I|$ collection requests, the
ground-truth pass issues $|I| \cdot m$ item requests, and the cross-identity pass
issues at most $|I| \cdot (|I|-1) \cdot m$ item requests. The total is therefore
$O(|I|^2 \cdot m)$ requests. Since a tester typically configures a small, fixed
number of identities, the request count is linear in $m$, which is the sample size
the operator chooses rather than the size of the target's store.

The confirmation predicate makes the tool conservative in a useful direction.
A finding is raised only when the attacker's response carries a success status and
an object whose identifier equals the requested one. If that condition holds, the
service genuinely returned the victim's object to an unauthorized caller, so a
raised finding corresponds to a real leak under the ownership definition of Section
\ref{sec:threat}. In this sense the method is sound with respect to that
definition, and the empty result on the hardened target in Section \ref{sec:eval}
is consistent with that property. The method is not complete. It can only test
objects it discovers, so a service that exposes no listing and uses unguessable
identifiers yields little to test, and it can miss a leak when the victim's data is
returned in a transformed body that does not echo the requested identifier. Both
gaps are visibility limits rather than logical errors, and both are addressable by
supplying explicit seeds or a custom body matcher in configuration. The design
deliberately trades some completeness for a low false-positive rate, because a
scanner that cries wolf is quickly ignored inside a build pipeline.

\section{Implementation}
\label{sec:impl}
AuthProbe is implemented in Python. The scanner core depends only on a small HTTP
client and a YAML parser, which keeps installation light and portable. The tool
parses the OpenAPI document, performs the algorithms of Section \ref{sec:method},
and renders reports in four formats. The console format gives a human summary, the
JSON format is intended for programmatic consumption, the Markdown format is
convenient for pull-request comments, and the JUnit format lets an existing test
dashboard display each finding as a failed test case. A command-line interface
exposes a single \code{scan} subcommand that reads a configuration file, selects
output formats, writes reports to a directory, and sets the exit code.

To make the tool testable and demonstrable without any external system, the
project ships two target services built with a standard Python web framework. The
vulnerable target reproduces the McHire failure class: it issues sequential
integer identifiers and its item endpoint authenticates the caller but omits the
ownership check, so any authenticated user can read any application. The hardened
target is the corrected counterpart: it issues non-sequential universally unique
identifiers, enforces a deny-by-default ownership check at the point of access,
and returns a uniform not-found status for both unauthorized and missing objects.
Both services expose an identical contract, so a single configuration scans either
one, which makes the pair a clean before-and-after demonstration.

Two engineering choices support responsible use. First, the tool prints an
authorized-use banner on every run. Second, it refuses to scan a target whose host
is not local unless the operator passes an explicit acknowledgment flag or sets an
allow-remote option in configuration. These measures do not prevent misuse by a
determined operator, but they make the intended scope explicit and reduce the risk
of an accidental scan of an unintended host.

\subsection{Configuration}
A scan is described by a short configuration file, shown in Listing
\ref{lst:config}. It names the target, points at the specification, and lists the
identities the operator controls, each as a set of headers. Optional settings
select the severity threshold, the enumeration radius, and whether remote targets
are permitted. Resources are detected automatically from the specification, and
the file may add explicit resource definitions when a contract does not follow the
usual collection and item convention.

\begin{lstlisting}[caption={A minimal AuthProbe configuration.},label=lst:config]
target:
  base_url: "http://127.0.0.1:8000"
  spec: "auto"
identities:
  - name: alice
    headers: { Authorization: "Bearer alice-token" }
  - name: bob
    headers: { Authorization: "Bearer bob-token" }
settings:
  fail_on: high
\end{lstlisting}

\subsection{Probes and continuous integration}
Table \ref{tab:probes} summarizes the four probes, the OWASP category each maps
to, and the signal that triggers a finding. In a pipeline, AuthProbe is invoked as
a single step that starts the service, runs the scan, and inspects the exit code,
as shown in Listing \ref{lst:ci}. When any finding reaches the configured
threshold the step fails, so a regression that reintroduces an object-level
authorization flaw blocks the change before release. The JUnit output can be
attached to the build so that each finding appears as a failed test case in the
existing dashboard.

\begin{table}[t]
\caption{AuthProbe probes and their trigger signals.}
\label{tab:probes}
\centering
\footnotesize
\begin{tabular}{@{}llll@{}}
\toprule
Probe & OWASP & Sev. & Trigger signal \\
\midrule
BOLA            & API1:2023 & high & victim object returned to attacker \\
IDOR walk       & API1:2023 & high & non-owned object reached by stepping id \\
Enumerable id   & API1:2023 & med  & all identifiers are numeric \\
Missing auth    & API2:2023 & crit & data returned with no credentials \\
Existence oracle& API1:2023 & low  & forbidden and missing differ in status \\
\bottomrule
\end{tabular}
\end{table}

\begin{lstlisting}[caption={AuthProbe as a continuous integration gate.},label=lst:ci]
- name: AuthProbe
  run: |
    authprobe scan --config authprobe.yaml \
      --format junit --out out/ --fail-on high
\end{lstlisting}

\subsection{Extensibility}
The implementation separates concerns so that new capability can be added without
disturbing the core. Each probe is an independent function that receives the
discovered ownership map and returns findings, so a new check, for example one that
tests write verbs, is a self-contained addition. Reporters are likewise pluggable,
which is how the same result renders to console, JSON, Markdown, and JUnit, and how
a further format such as SARIF could be added for code-scanning dashboards.
Identity construction is a small adapter as well, so establishing a session through
a login exchange rather than a static header is a localized change. This structure
keeps the tool small while leaving clear seams for the extensions outlined in
Section \ref{sec:conc}.

\section{Evaluation}
\label{sec:eval}
We evaluate three questions. Does AuthProbe detect the planted McHire-class flaw?
Does it avoid false positives on a hardened service? How does its running time
scale with the size of the target?

\subsection{Experimental setup}
All experiments run against the two shipped target services on a single host, with
two tester identities named alice and bob. For the detection and false-positive
experiments each identity creates three applications. For the scaling experiment
each identity creates a varying number of applications, from one to fifty. The
targets store data in memory and are reset before each run so that results are
reproducible. Because the targets are synthetic and contain only fabricated data,
the evaluation involves no real personal information.

\subsection{Detection and false positives}
Table \ref{tab:findings} summarizes the outcome. On the vulnerable target with
three applications per identity, AuthProbe reports seven high-severity findings
and one medium-severity finding. The high-severity findings comprise the confirmed
cross-identity reads in both directions, that is alice reading each of bob's
objects and bob reading each of alice's, together with a successful enumeration
walk, and the medium-severity finding is the enumerable-identifier scheme. On the
hardened target, AuthProbe reports nothing. The tool therefore detects the planted
vulnerability class in full and produces no findings against the corrected
service, which is the behavior a build gate requires.

\begin{table}[t]
\caption{Findings by severity on the two targets, three applications per identity.}
\label{tab:findings}
\centering
\begin{tabular}{@{}lcccc@{}}
\toprule
Target & Critical & High & Medium & Low \\
\midrule
Vulnerable & 0 & 7 & 1 & 0 \\
Hardened   & 0 & 0 & 0 & 0 \\
\bottomrule
\end{tabular}
\end{table}

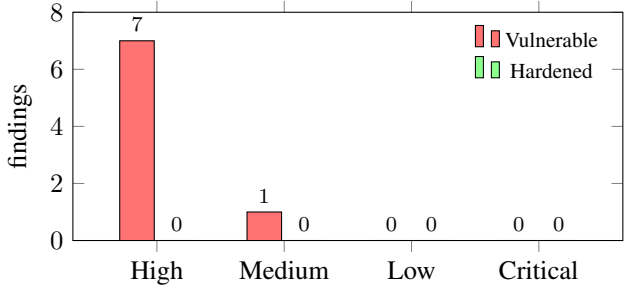
\begin{figure}[t]
\centering
\begin{tikzpicture}
\begin{axis}[
  ybar, width=\columnwidth, height=4.6cm, bar width=13pt,
  ymin=0, ymax=8, ylabel={findings}, ylabel near ticks,
  symbolic x coords={High,Medium,Low,Critical},
  xtick=data, enlarge x limits=0.22,
  legend style={at={(0.98,0.97)}, anchor=north east, font=\footnotesize, draw=none},
  nodes near coords, nodes near coords style={font=\footnotesize},
]
\addplot[fill=red!55] coordinates {(High,7)(Medium,1)(Low,0)(Critical,0)};
\addplot[fill=green!45] coordinates {(High,0)(Medium,0)(Low,0)(Critical,0)};
\legend{Vulnerable, Hardened}
\end{axis}
\end{tikzpicture}
\caption{Findings by severity. The vulnerable target yields seven high and one
medium finding, while the hardened target yields none.}
\label{fig:severity}
\end{figure}

Figure \ref{fig:severity} shows the same result as a chart. The separation is
categorical rather than marginal: the vulnerable target produces high-severity
findings, and the hardened target produces none, so a threshold at the high level
cleanly distinguishes the two.

\subsection{Scalability}
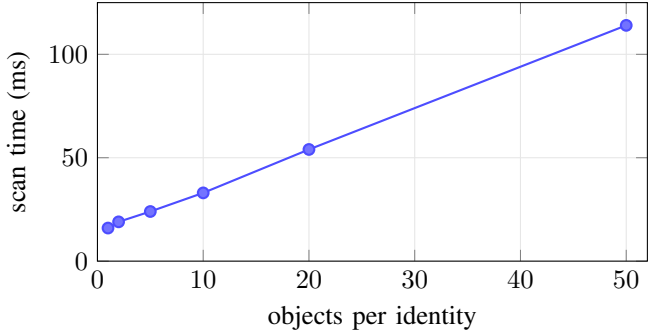
\begin{figure}[t]
\centering
\begin{tikzpicture}
\begin{axis}[
  width=\columnwidth, height=5.0cm,
  xlabel={objects per identity}, ylabel={scan time (ms)},
  ylabel near ticks, xmin=0, xmax=52, ymin=0, ymax=125,
  grid=both, grid style={gray!20},
  mark options={fill=blue!55},
]
\addplot[blue!70, thick, mark=*] coordinates {
  (1,16)(2,19)(5,24)(10,33)(20,54)(50,114)
};
\end{axis}
\end{tikzpicture}
\caption{Scan time against the number of objects per identity on the vulnerable
target. The trend is close to linear, consistent with the cost analysis when the
number of tester identities is small and fixed.}
\label{fig:scale}
\end{figure}

Figure \ref{fig:scale} plots the running time of a full scan against the number of
applications per identity, from one to fifty, on the vulnerable target. The
measured times grow from sixteen milliseconds at a single object per identity to
one hundred fourteen milliseconds at fifty, and the growth is close to linear, as
predicted by the cost analysis of Algorithm \ref{alg:bola} with a fixed, small
number of identities. Over the same range the number of findings grows from four
to one hundred two, because the tool reports one confirmed read per victim object
per attacker in addition to the fixed enumeration and enumerable-identifier
findings. The tool therefore remains fast on realistically sized collections when
seeded with a bounded sample per identity, and the linear trend means the cost is
governed by the sample size the operator chooses rather than by the total size of
the target's data store.

\begin{table}[t]
\caption{Raw scalability measurements on the vulnerable target.}
\label{tab:scale}
\centering
\footnotesize
\begin{tabular}{@{}rrr@{}}
\toprule
Objects per identity & Scan time (ms) & Findings \\
\midrule
1  & 16  & 4   \\
2  & 19  & 6   \\
5  & 24  & 12  \\
10 & 33  & 22  \\
20 & 54  & 42  \\
50 & 114 & 102 \\
\bottomrule
\end{tabular}
\end{table}

\subsection{Interpretation}
The evaluation supports the central claim. AuthProbe detects a BOLA and IDOR
pattern that a signature-based firewall and a single-identity scanner would miss,
it does so with a confirmation predicate that produced no false positives on the
hardened target, and it runs quickly enough to sit inside a routine build. The
result is modest in scope by design, since the targets are synthetic and the
purpose of this evaluation is to establish correctness of the mechanism rather
than a field study of prevalence. A larger study against real, authorized targets
is future work and is discussed below.

\subsection{Case study: reproducing the McHire failure chain}
The vulnerable target reproduces the two defects that combined in the motivating
incident. Its item endpoint authenticates the caller but omits the ownership
check, and it issues sequential integer identifiers. When alice and bob each
create an application, alice owns identifier one and bob owns identifier two. The
BOLA probe fetches bob's object while authenticated as alice, receives a success
status and a body whose identifier equals two, and raises the finding shown in
Listing \ref{lst:finding}. The enumeration probe independently starts from alice's
identifier one, steps to two, reaches bob's object, and raises an IDOR finding,
which mirrors the decrementing walk used in the real incident. The enumerable
identifier scheme itself is reported at medium severity because numeric identifiers
are walkable even where a check is currently present. The hardened target defeats
all three findings at once by enforcing ownership, returning a uniform not-found
status, and issuing unguessable identifiers.

\begin{lstlisting}[caption={A confirmed BOLA finding in JSON form.},label=lst:finding]
{
  "probe": "bola",
  "severity": "high",
  "resource": "applications",
  "endpoint": "/applications/{app_id}",
  "evidence": {"attacker": "alice", "victim": "bob",
               "object_id": 2, "status": 200}
}
\end{lstlisting}

\subsection{Comparison with conventional defenses}
It is worth stating plainly why the defenses most teams already run do not catch
this class. A signature-based web application firewall inspects a request for known
malicious patterns, yet the attacker's request here is a well-formed fetch with a
valid session and an ordinary identifier, so there is no signature to match. A
conventional dynamic scanner that operates under a single identity can exercise the
endpoint and confirm that it returns data, but with only one identity it has no way
to know that the returned object belongs to someone else, so it cannot label the
response as a leak. AuthProbe succeeds precisely because it holds the second
identity and the ownership map, which lets it recognize a cross-identity read for
what it is. The requirement for two identities is therefore not an inconvenience
but the very feature that makes the flaw observable.

\subsection{Why the hardened target yields no findings}
It is instructive to trace each probe against the hardened target, because the
empty result is not an accident of configuration but a direct consequence of three
corrections. The BOLA probe fetches a victim object as the attacker, but the
service compares the object's owner against the caller and returns a not-found
status, so the leak predicate is never satisfied. The enumeration probe cannot even
begin its walk in a meaningful way, because the identifiers are universally unique
values rather than integers, so there is no neighbor to step to and the numeric
check that flags enumerable identifiers does not fire. The missing-authentication
probe receives an unauthorized status for every credential-free request, and the
existence-oracle probe observes the same not-found status for an unauthorized object
and for a missing one, so no oracle is reported. Each probe is defeated by the
specific control it is designed to exercise, which is the behavior a faithful
detector should exhibit and which gives confidence that a green result is
meaningful rather than a blind spot.

\subsection{Request volume}
The scan cost is easiest to reason about in terms of request volume rather than
wall-clock time, since the latter depends on the environment. For the reported
setup with two identities and $m$ objects each, ownership discovery issues two
collection requests, the ground-truth pass issues $2m$ item requests, the
cross-identity pass issues $2m$ item requests, and the enumeration probe issues a
bounded number proportional to its radius. The dominant term is therefore
$4m$ item requests, which matches the near-linear timing curve of Figure
\ref{fig:scale} and confirms that an operator controls the cost directly through
the sample size configured per identity. A larger sample increases confidence that
a flaw would be caught while keeping the scan within a predictable budget.

\section{Discussion and Limitations}
\label{sec:disc}
AuthProbe is black box by design, and that choice brings both its strengths and
its limits. Because it needs only the specification and network access, it applies
to services whose source code is unavailable and it fits naturally into deployment
pipelines, which is where static analyzers such as BolaRay cannot reach. The cost
of the black-box stance is that the tool can only reason about objects it can name.
It therefore depends on the collection operation to discover ownership, or on
explicit seeds in configuration when no suitable collection operation exists. An
API that exposes no listing and uses unguessable identifiers will yield few
objects to test, which is a limitation of visibility rather than a false negative
in the usual sense.

The present version focuses on read access through the item operation, which is
the dominant BOLA pattern and the one behind the motivating incident. Write verbs
such as update and delete can carry the same flaw, and extending the probes to
those verbs is a natural next step that reuses the same ownership model. AuthProbe
also targets object-level authorization specifically. Function-level
authorization, the fifth item on the OWASP list, concerns whether a caller may
invoke an operation at all, and detecting it requires a notion of roles that the
current tool does not model. Finally, the confirmation predicate keeps false
positives low, but it can miss a leak when a service returns the victim's data in
a shape that does not carry the requested identifier, for example a wrapped or
transformed body. The predicate is configurable to accommodate such shapes, but a
default deployment may under-report in those cases, which is a conservative
failure mode.

Taken together these limitations position AuthProbe as a complement to, rather
than a replacement for, source-level analysis and manual review. Its value is that
it turns the most common and most damaging API flaw into a repeatable,
evidence-producing check that a team can run on every build without specialist
tooling. Used together with a source-level analyzer during development and a manual
review before a major release, it closes the deployment-time gap in which a flaw
can slip into a running service unnoticed. The three layers reinforce one another:
static analysis reasons about code that may never ship, manual review brings human
judgment to complex cases, and AuthProbe checks the artifact that is actually
deployed.

\subsection{Threats to validity}
Several factors bound the strength of the evaluation. The construct we measure is
whether a returned object belongs to a principal that did not own it, which we
operationalize through the ownership definition of Section \ref{sec:threat}; a
target whose collection operation does not faithfully reflect ownership could
distort that construct, although such a discrepancy would itself be a finding worth
surfacing. Internally, the targets are in-memory services on a single host, so the
absolute timings exclude network latency and would be larger against a remote
service; the linear trend, however, follows from the request-count analysis and
does not depend on the absolute constants. Externally, the two targets are
synthetic and were authored to exhibit and to fix the studied flaw, so the results
establish that the mechanism works rather than how common the flaw is in the field.
Measuring prevalence requires an authorized study across real deployments, which we
identify as future work. We mitigate author bias in the target design by keeping
the hardened target a faithful minimal correction of the vulnerable one, changing
only the identifier scheme, the ownership check, and the error uniformity.

\subsection{Deployment models}
AuthProbe fits three deployment models. In the pipeline model it runs on every
build against an ephemeral instance of the service, which catches regressions
early and keeps the check close to the code change that caused it. In the staging
model it runs on a schedule against a pre-production environment seeded with
synthetic accounts, which exercises a more realistic configuration including
gateways and reverse proxies. In the assessment model an authorized tester runs it
once against a production or production-like system as part of a broader review,
using the explicit acknowledgment flag. The same configuration and the same probes
serve all three, and only the cadence and the target environment change.

\section{Ethical and Legal Considerations}
\label{sec:ethics}
A tool that reads one user's data as another user is only ethical when the
operator is authorized to test the target. AuthProbe is built for that setting.
The bundled vulnerable service exists solely so that the tool can be validated
against a system the operator controls, and it is clearly marked as unfit for
deployment. The runtime guardrail described in Section \ref{sec:impl} refuses
non-local targets without an explicit acknowledgment, which reduces the chance of
an accidental scan and records the operator's assertion of authorization.
Operators remain responsible for compliance with applicable law, including
computer-misuse statutes, and for coordinated disclosure when a finding concerns a
third party. Because recruitment data is personal data, testing against real
systems also engages data-protection obligations such as those in the General Data
Protection Regulation \cite{gdpr}, and operators should prefer synthetic or
consented data when validating a deployment. The project itself ships only
fabricated data and encourages the same practice downstream. When AuthProbe is used
against a system operated by another party under authorization, any confirmed
finding should be handled through coordinated disclosure. In practice this means
reporting the specific endpoint, the ownership violation, and the minimal evidence
needed to reproduce it, allowing the operator a reasonable window to remediate
before any public description, and never retaining or redistributing the personal
data that a leak exposes. The tool supports this workflow by recording, for each
finding, only the identifiers and status codes required to reproduce it rather than
the full contents of a victim record, which keeps the evidence actionable while
minimizing the sensitive data it captures.

\section{Conclusion and Future Work}
\label{sec:conc}
Broken Object-Level Authorization remains the most common and most damaging flaw
in the APIs that handle applicant data, and it is invisible to the defenses that
most teams already run. AuthProbe addresses that gap with a black-box,
specification-driven scanner that detects the flaw by acting under multiple
identities and confirming a leak through response differencing, and that fits a
delivery pipeline through a severity-thresholded exit code and machine-readable
reports. On a testbed that reproduces the McHire failure class the tool detects
every planted cross-identity read, produces no findings on the hardened
counterpart, and scales linearly with the size of the sample under test.

Future work proceeds along several lines that reuse the present design.
\begin{itemize}
\item \textbf{Write-verb and nested-resource coverage.} Extending the probes to
update and delete operations, and to nested resources such as an application under
a requisition, broadens coverage while reusing the same ownership model, since the
leak predicate and the discovery step carry over unchanged.
\item \textbf{Function-level authorization.} Adding a notion of roles enables
detection of broken function-level authorization, the fifth item on the OWASP list,
which asks whether a caller may invoke an operation at all rather than whether it
may touch a specific object.
\item \textbf{Richer identity establishment.} A login-flow adapter would let an
identity be established through a credential exchange, extracting a token from a
sign-in response, which removes the need to pre-provision static headers and better
mirrors real client behavior.
\item \textbf{Standardized reporting.} Emitting results in the Static Analysis
Results Interchange Format would let findings appear directly in code-scanning
dashboards alongside other security results, improving triage.
\item \textbf{An authorized field study.} Measuring the prevalence of the flaw
across real, authorized recruitment APIs, and comparing the black-box detection
rate against source-level tools, would move the work from a correctness
demonstration to an empirical characterization of the problem in the wild.
\end{itemize} Beyond the specific tool, the broader aim is to make object-level authorization a
property that teams verify by default rather than discover after a breach. The
history of software security suggests that a class of flaw recedes only when
checking for it becomes routine, cheap, and automatic, as happened with memory
safety scanners and dependency auditing. Broken Object-Level Authorization is
overdue for the same treatment, and a small, specification-driven, pipeline-native
check is a step toward that routine. The implementation, the demonstration targets,
and the experimental scripts are available under the Apache 2.0 license at the
address on the title page.

\end{document}